\documentclass[conference]{IEEEtran}
\IEEEoverridecommandlockouts
\usepackage{cite}
\usepackage{amsmath,amssymb,amsfonts,algorithm,algorithmic,graphicx,epstopdf}
\usepackage{algorithmic}
\usepackage{graphicx}
\usepackage{textcomp}
\usepackage{xcolor}
\usepackage{subfloat}
\def\BibTeX{{\rm B\kern-.05em{\sc i\kern-.025em b}\kern-.08em
    T\kern-.1667em\lower.7ex\hbox{E}\kern-.125emX}}
\usepackage[left=0.7in,right=0.7in,top=0.7in,bottom=1.1in]{geometry}
\begin{document}

\title{High-Precision Channel Estimation for Sub-Noise Self-Interference Cancellation
\thanks{This work was supported in part by the National Key R\&D Program of China under Grant 2020YFB1807802, and in part by National Natural Science Foundation of China under Grant 62171006.}
}

\author{\IEEEauthorblockN{Dongsheng Zheng, Lifeng Lin, Wenyao Li, Bingli Jiao}
\IEEEauthorblockA{School of Electronics, Peking University, Beijing, China \\
e-mail: \{zhengds, linlifeng\}@pku.edu.cn, liwenyao@stu.pku.edu.cn, jiaobl@pku.edu.cn}
}

\maketitle

\begin{abstract}
Self-interference cancellation plays a crucial role in achieving reliable full-duplex communications. In general, it is essential to cancel the self-interference signal below the thermal noise level, which necessitates accurate reconstruction of the self-interference signal. In this paper, we propose a high-precision channel estimation method specifically designed for sub-noise self-interference cancellation. Exploiting the fact that all transmitted symbols are known to their respective receivers, our method utilizes all transmitted symbols for self-interference channel estimation. Through analytical derivations and numerical simulations, we validate the effectiveness of the proposed method. The results demonstrate the superior performance of our approach in achieving sub-noise self-interference cancellation.
\end{abstract}


\section{Introduction}
The full duplex technique has emerged as a key technology driving the progression from 5G to 6G for its potential to double the spectrum efficiency \cite{Giordani_M2020,Alsabah_M2021}. However, this approach encounters a major challenge: self-interference from the local transmitter while receiving the desired signal from a remote device. Consequently, effective self-interference cancellation (SIC) is vital for the successful implementation of full duplex communications \cite{Xin_Y2015,Ma_M2020}. SIC techniques can be sequentially implemented in three domains: the antenna domain, analog domain, and digital domain \cite{Kolodziej_K2019}. In this paper, our primary focus is on digital self-interference cancellation, which leverages transmitted baseband symbols and estimated channel coefficients to reconstruct the self-interference signal. This reconstructed self-interfering signal can then be subtracted from the received samples, facilitating the suppression of self-interference. Therefore, achieving high-precision self-interference channel estimation is crucial for the accurate reconstruction of the self-interference signal and effectively canceling interference below the noise level.

Coherent detection often relies on a pilot-based frame structure to estimate channel coefficients. Similarly, in the case of self-interference channel estimation, pilot symbols are used to calculate the self-interference coefficients \cite{Kong_C2019}. However, to achieve more accurate estimation, a larger number of pilot symbols need to be inserted into data frames, resulting in reduced spectrum efficiency. Additionally, the self-interference channel estimation is susceptible to interference from the desired signal. To address the need for precise channel estimation, Li et al. \cite{Li_T2021} proposed a novel frame structure in which the desired signal remains silent during the pilot symbol interval of the self-interference signal. However, this method requires full-duplex communication transceivers to be synchronized, thereby increasing the complexity of system design. Moreover, the inclusion of silent symbols within data frames further reduces spectrum efficiency.

It is worth noting that relying solely on pilot symbols for self-interference channel estimation disregards the fact that all transmit symbols, including both pilot and data symbols, are known for estimating the interference channel in full-duplex communications. Considering this important characteristic of the self-interference signal, this paper presents a novel and precise method for estimating the interference channel. The proposed method eliminates the requirement for additional pilot symbols and enables the utilization of all known transmitted symbols for self-interference channel estimation, thereby substantially improving estimation accuracy. With the availability of high-precision estimated channel coefficients, the self-interference can be effectively suppressed below the noise level.

\section{Self-Interference Cancellation below Noise}
In this section, we provide a comprehensive elaboration on the proposed digital self-interference cancellation method. Initially, we introduce the innovative high-precision self-interference channel estimation approach, which serves as a fundamental component of our method. Building upon this, we subsequently introduce the technique for achieving digital self-interference cancellation below the noise level.

\subsection{High-Precision Interference Channel Estimation}
The accurate reconstruction of self-interference in digital domain necessitates high-precision interference channel estimation. Based on the fact that all transmitted symbols are known by the transceiver, we propose a method for estimating self-interference channel coefficients with high precision. To be specific, the proposed method utilizes all transmitted symbols rather than just pilot symbols to estimate channel coefficients, which leads to a reduction in the error of channel estimation. The details are stated as follows.

Denote the transmitted symbols by $\mathbf{x}=[x_1,x_2,\cdots,x_{N-1},x_N]^{\mathrm{T}}$, where $N$ is the length of the transmitted symbol sequence and ${(\cdot)}^{\mathrm{T}}$ is the transpose operator. Correspondingly, the received symbols are denoted by $\mathbf{y}=[y_1, y_2, \cdots, y_{N-1}, y_N]^{\mathrm{T}}$. The self-interference channel is characterized by the channel impulse response, where the coefficient of $k$-th tape is represented by $h_k$. Here, we define the column vector $\mathbf{h}$ as $\mathbf{h}=[h_1,h_2,\cdots,h_{K-1},h_K]^{\mathrm{T}}$ with $K$ denoting the tapes of channel impulse response. Then, we have
\begin{equation}\label{rec_conv}
	\mathbf{y}=\mathbf{x}*\mathbf{h}+\boldsymbol{\omega},
\end{equation}
where $*$ denotes the convolution operation. The column vector $\boldsymbol{\omega}=[\omega_1,\omega_2,\cdots,\omega_{N-1},\omega_{N}]^T$ represents the additive noise component, where each entry is independent and identically distributed (i.i.d.) with zero mean and variance of $\sigma_{\omega}^2$. It is noted here that the noise component $\boldsymbol{\omega}$ encompasses the combined effects of both the desired signal and the additive white Gaussian noise. For the sake of simplicity in derivation, we can rewrite Eq. (\ref{rec_conv}) in its matrix formation as follows,
\begin{align}\label{rec_matrix}
	\mathbf{y}&=\left[\begin{array}{ccccc}
		x_1 & 0 & \cdots & 0 & 0 \\
		x_2 & x_1 & \cdots & 0 & 0 \\
		\vdots & \vdots & \ddots & \vdots & \vdots \\
		x_{N-1} & x_{N-2} & \cdots & x_{N-K+1} & x_{N-K} \\
		x_N & x_{N-1} & \cdots & x_{N-K+2} & x_{N-K+1}
	\end{array}\right] \mathbf{h}+\boldsymbol{\omega} \nonumber\\
	&\triangleq \mathbf{X}\mathbf{h}+\boldsymbol{\omega}.
\end{align}

The least-square (LS) estimation metric is used to estimate the channel coefficients vector $\mathbf{h}$. Specifically, the estimated value $\hat{\mathbf{h}}$ can be obtained by solving the following optimization problem.
\begin{equation}
	O(\hat{\mathbf{ h}})=\mathop {\min }\limits_{{\hat{\mathbf{ h}}}} \left\| {\mathbf{y}} - {\mathbf{X}\hat{\mathbf{h}}} \right\|_F^2,
\end{equation}
where $\|\cdot\|_{F}$ denotes the Frobenius norm.

The first-order derivation of the objective function $O(\hat{\mathbf{ h}})$ is formulated as
\begin{align}
	O'(\hat{\mathbf{ h}})=&\frac{{\partial \left\| {{\mathbf{y}} - {\mathbf{X}\hat{\mathbf{h}}}} \right\|_F^2}}{{\partial {\hat{\mathbf{h}}}}} = \frac{{\partial \left[ {{{({\mathbf{y}} - {\mathbf{X}\hat{\mathbf{h}}})}^H}({\mathbf{y}} - {\mathbf{X}\hat{\mathbf{h}}})} \right]}}{{\partial {\hat{\mathbf{h}}}}} \nonumber\\ 
	&= \frac{{\partial \left\{ {{{\mathbf{y}}^H}{\mathbf{y}} + {{\hat{\mathbf{h}}}^H}{{\mathbf{X}}^H}{\mathbf{X}\hat{\mathbf{h}}} - {{\mathbf{y}}^H}{\mathbf{X}\hat{\mathbf{h}}} - {{\hat{\mathbf{h}}}^H}{{\mathbf{X}}^H}{\mathbf{y}}} \right\}}}{{\partial {\hat{\mathbf{h}}}}} \nonumber\\
	&= 2{{\mathbf{X}}^H}{\mathbf{X}\hat{\mathbf{h}}} - 2{{\mathbf{X}}^H}{\mathbf{y}},
\end{align}
where $(\cdot)^H$ denotes the conjugate transpose of a matrix. The estimated value $\hat{\mathbf{h}}$ can be obtained by solving the equation $O'(\hat{\mathbf{ h}})=0$, which is calculated by
\begin{equation}\label{h_esti}
	\hat{\mathbf{h}}=\left({\mathbf{X}}^H\mathbf{X}\right)^{-1}{{\mathbf{X}}^H}{\mathbf{y}}\triangleq \mathbf{X}^{\dagger}{\mathbf{y}},
\end{equation}
where $\mathbf{X}^{\dagger}=\left({\mathbf{X}}^H\mathbf{X}\right)^{-1}{{\mathbf{X}}^H}$ is the Moore-Penrose inverse of the matrix $\mathbf{X}$. Substituting Eq. (\ref{rec_matrix}) into Eq. (\ref{h_esti}), we can obtain
\begin{equation}\label{h_esti2}
	\hat{\mathbf{h}}=\mathbf{h} + \mathbf{X}^{\dagger}\boldsymbol{\omega}.
\end{equation}

Then, the mean square error of channel estimation is calculated by
\begin{align}\label{mse_h}
	\sigma_E^2 &= \frac{1}{K}\sum\limits_{k=1}^{K}\mathbb{E}\left\{|\hat{h}_k-{h}_k|^2\right\}\nonumber\\
	&=\frac{1}{K}\mathrm{Tr}\left\{\mathbb{E}\left\{(\hat{\mathbf{h}}-{\mathbf{h}})(\hat{\mathbf{h}}-{\mathbf{h}})^H\right\}\right\}\nonumber\\
	&=\frac{1}{K}\mathrm{Tr}\left\{\mathbb{E}\left\{\mathbf{X}^{\dagger}\boldsymbol{\omega}\boldsymbol{\omega}^H(\mathbf{X}^{\dagger})^H\right\}\right\}\nonumber\\
	&=\frac{\sigma_{\omega}^2}{K}\mathrm{Tr}\left\{\mathbf{X}^{\dagger}(\mathbf{X}^{\dagger})^H\right\}=\frac{\sigma_{\omega}^2}{K}\mathrm{Tr}\left\{\left({\mathbf{X}}^H\mathbf{X}\right)^{-1}\right\}.
\end{align}

\begin{figure}[!t]
	\includegraphics[width=0.48\textwidth]{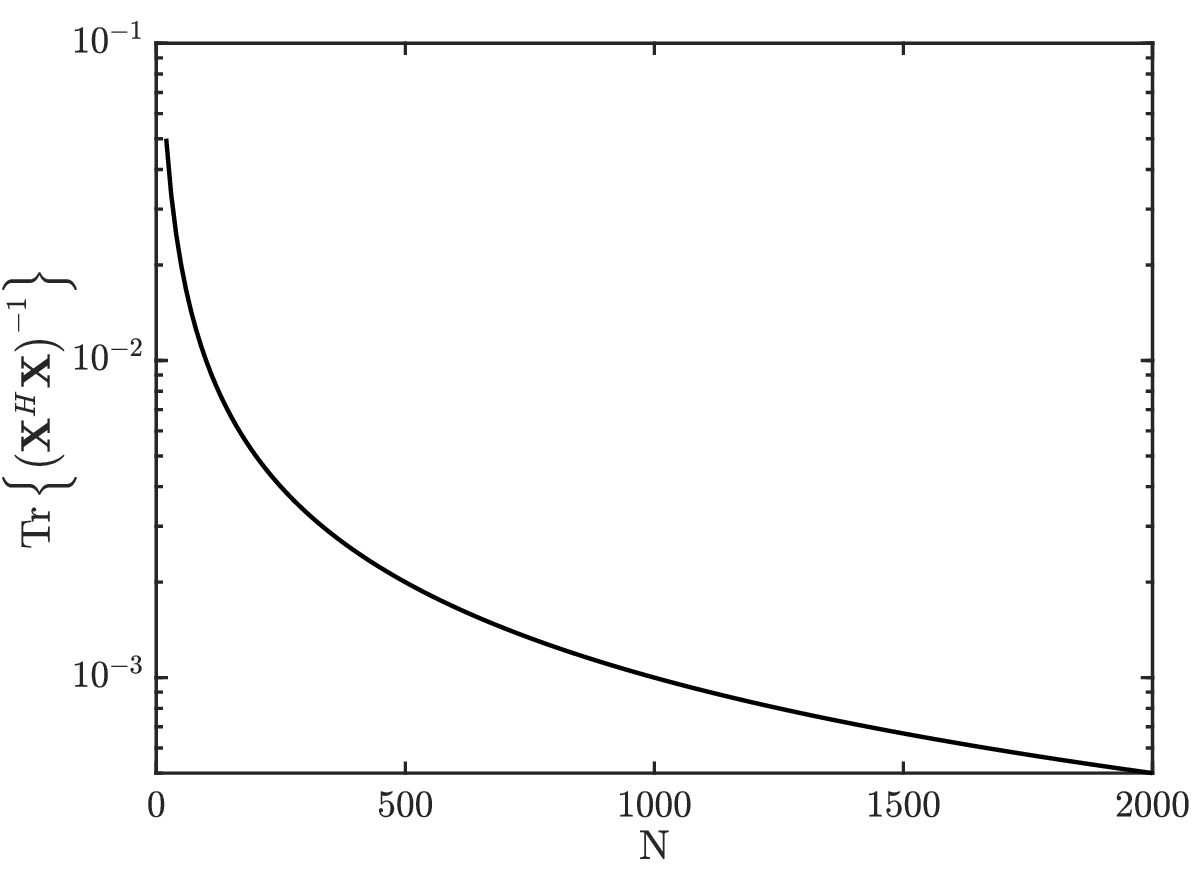}
	\caption{The value of $\mathrm{Tr}\left\{({\mathbf{X}}^H\mathbf{X})^{-1}\right\}$ versus the length of the transmitted symbol sequence $N$.}
	\label{trace_figure}
\end{figure}

To achieve high-precision channel estimation and reduce the mean square error $\sigma_E^2$, two schemes can be implemented based on Eq. (\ref{mse_h}). On the one hand, the value of $\sigma_{\omega}^2$ can be decreased by reducing the interference at pilot symbols. Towards this end, it is necessary for the desired signal to maintain silent during the pilot symbol duration of the self-interference signal \cite{Li_T2021}. However, it increases the complexity of system design and decreases the spectral efficiency gain of full duplex communications. On the other hand, the value of $\mathrm{Tr}\left\{({\mathbf{X}}^H\mathbf{X})^{-1}\right\}$ can be decreased by increasing the number of transmitted symbols employed for self-interference channel estimation. To substantiate this assertion, Fig. \ref{trace_figure} illustrates the trend of the value  $\mathrm{Tr}\left\{({\mathbf{X}}^H\mathbf{X})^{-1}\right\}$ with respect to the number of transmitted symbols $N$, where the taps of channel impulse response $K$ is set to 20, and the transmitted symbol sequence $\mathbf{x}$ are modulated using binary phase-shift keying (BPSK). It can be inferred from Fig. \ref{trace_figure} that the mean square error $\sigma_E^2$ tends to zero as the number of symbols approaches infinity. 

 Building upon this observation, we propose a high-precision self-interference channel estimation method, that is, using all transmitted symbols to estimate channel coefficient vector $\mathbf{ h}$. To elucidate the operation of the proposed high-precision channel estimation method, we will consider an illustrative example involving a self-interference channel characterized by only a single-tap impulse response. In this scenario, according to Eq. (\ref{rec_matrix}), the received signal can be simplified to
\begin{equation}
	\mathbf{y}= h_1\mathbf{x}+\boldsymbol{\omega},
\end{equation}
where $h_1$ represents the single-tap channel impulse response. Substituting $\mathbf{X}=\mathbf{x}$ into Eq. (\ref{h_esti}), we can obtain the estimated value of $h_1$ under the LS estimation metric as 
\begin{equation}\label{h1_esti}
	\hat{h}_1=\frac{\mathbf{x}^H\mathbf{y}}{\mathbf{x}^H\mathbf{x}}=\frac{\sum\limits_{n=1}^{N}x_n' y_n}{\sum\limits_{n=1}^{N}x_n'x_n}=h_1 + \frac{\sum\limits_{n=1}^{N}x_n' \omega_n}{NE_x},
\end{equation}
where $(\cdot)'$ denotes the conjugate of a variable, and $E_x$ is the symbol energy of transmitted symbols. It is noteworthy that the process of solving the estimated channel coefficient as shown in Eq. (\ref{h1_esti}) can be regarded as a despreading process. Meanwhile, referring to Eq. (\ref{mse_h}), the estimation mean square error is formulated as
\begin{equation}\label{mse_h1}
	\sigma_E^2=\frac{\sigma_{\omega}^2}{NE_x},
\end{equation}
which indicates that the estimation error of $h_1$ decreases as the number of transmitted symbols $N$ increases.

Through the presented illustrative example, we demonstrate the functionality of our proposed high-precision self-interference channel estimation. Subsequently, we will showcase how this high-precision channel estimation technique enhances the capability to cancel self-interference in full duplex communications.

\subsection{Digital Self-Interference Cancellation}
By combing the estimated self-interference channel coefficients $\hat{\mathbf{h}}$ and the transmitted symbols $\mathbf{x}$, we can reconstruct the received self-interference $\hat{\mathbf{r}}$ in digital domain. Specifically, the reconstructed signal is formulated as
\begin{equation}
	\hat{\mathbf{r}}=\mathbf{x}*\hat{\mathbf{h}}=\mathbf{X}\hat{\mathbf{h}}.
\end{equation}
Then, the residual self-interference $\Delta{\mathbf{r}}$ after digital-domain self-interference cancellation can be represented as 
\begin{equation}\label{RSI}
	\Delta{\mathbf{r}}= {\mathbf{y}} - \left(\hat{\mathbf{r}}+\boldsymbol{\omega}\right)=X(\mathbf{h}-\hat{\mathbf{h}}).
\end{equation}
Eq. (\ref{RSI}) suggests that the self-interference cancellation capability derives great benefit from the enhanced precision of self-interference channel estimation, namely, reduced estimation error $\Delta \mathbf{h} = \mathbf{h}-\hat{\mathbf{h}}$.

Furthermore, substituting Eq. (\ref{h_esti2}) into Eq. (\ref{RSI}), we have 
\begin{align}
	\Delta{\mathbf{r}} &= -\mathbf{X}\mathbf{X}^{\dagger}\boldsymbol{\omega}=\left[-\mathbf{X}\left({\mathbf{X}}^H\mathbf{X}\right)^{-1}{{\mathbf{X}}^H}\right]\boldsymbol{\omega}.
\end{align} 
Then, the variance of residual self-interference can be calculated by
\begin{align}\label{rsi_var}
	\sigma_{\mathrm{rsi}}^2&=\frac{1}{N}\sum\limits_{n=1}^{N}\mathbb{E}\left\{|\Delta r_n|^2\right\}=\frac{1}{N}\mathrm{Tr}\left\{\mathbb{E}\left\{\Delta{\mathbf{r}}\left(\Delta{\mathbf{r}}\right)^H\right\}\right\}\nonumber\\
	&=\frac{1}{N}\mathrm{Tr}\left\{\mathbb{E}\left\{\mathbf{X}\mathbf{X}^{\dagger}\boldsymbol{\omega}\boldsymbol{\omega}^H\left(\mathbf{X}^{\dagger}\right)^H\mathbf{X}^H\right\}\right\}\nonumber\\
	&=\frac{\sigma_{\omega}^2}{N}\mathrm{Tr}\left\{\mathbf{X}\mathbf{X}^{\dagger}\left(\mathbf{X}^{\dagger}\right)^H\mathbf{X}^H\right\}\nonumber\\
	&=\frac{\sigma_{\omega}^2}{N}\mathrm{Tr}\left\{\mathbf{X}\left({\mathbf{X}}^H\mathbf{X}\right)^{-1}{{\mathbf{X}}^H}\right\} \nonumber\\
	&=\frac{\sigma_{\omega}^2}{N}\mathrm{Tr}\left\{{{\mathbf{X}}^H}\mathbf{X}\left({\mathbf{X}}^H\mathbf{X}\right)^{-1}\right\}=\frac{K\sigma_{\omega}^2}{N}.
\end{align}
According to Eq. (\ref{rsi_var}), the effect of residual self-interference diminishes as the number of symbols utilized for estimating the self-interference channel increases. This observation validates the efficacy of our proposed high-precision channel estimation method in digital self-interference cancellation process. Furthermore, an optimistic outcome deduced from Eq. (\ref{rsi_var}) is that the self-interference can be suppressed below the noise level as the length of the transmitted symbol sequence $N$ surpasses a certain threshold. Theoretically, complete cancellation of self-interference is achievable as the value of $N$ approaches infinity.

\section{Numerical Results}
In this section, we evaluate the performance of the proposed sub-noise self-interference cancellation method, which is based on high-precision channel estimation. For the sake of simplicity and without sacrificing generality, we consider the channel impulse response $\mathbf{h}=\{h_1,h_2,\cdots,h_{K-1},h_K\}$ to be i.i.d. complex Gaussian random variables with zero mean and unit variance. Unless otherwise stated, the default self-interference to desired signal ratio is set to 10dB, and the default desired signal
 to additive white Gaussian noise ratrio is set to 20dB.

\begin{figure}[!t]
	\includegraphics[width=0.48\textwidth]{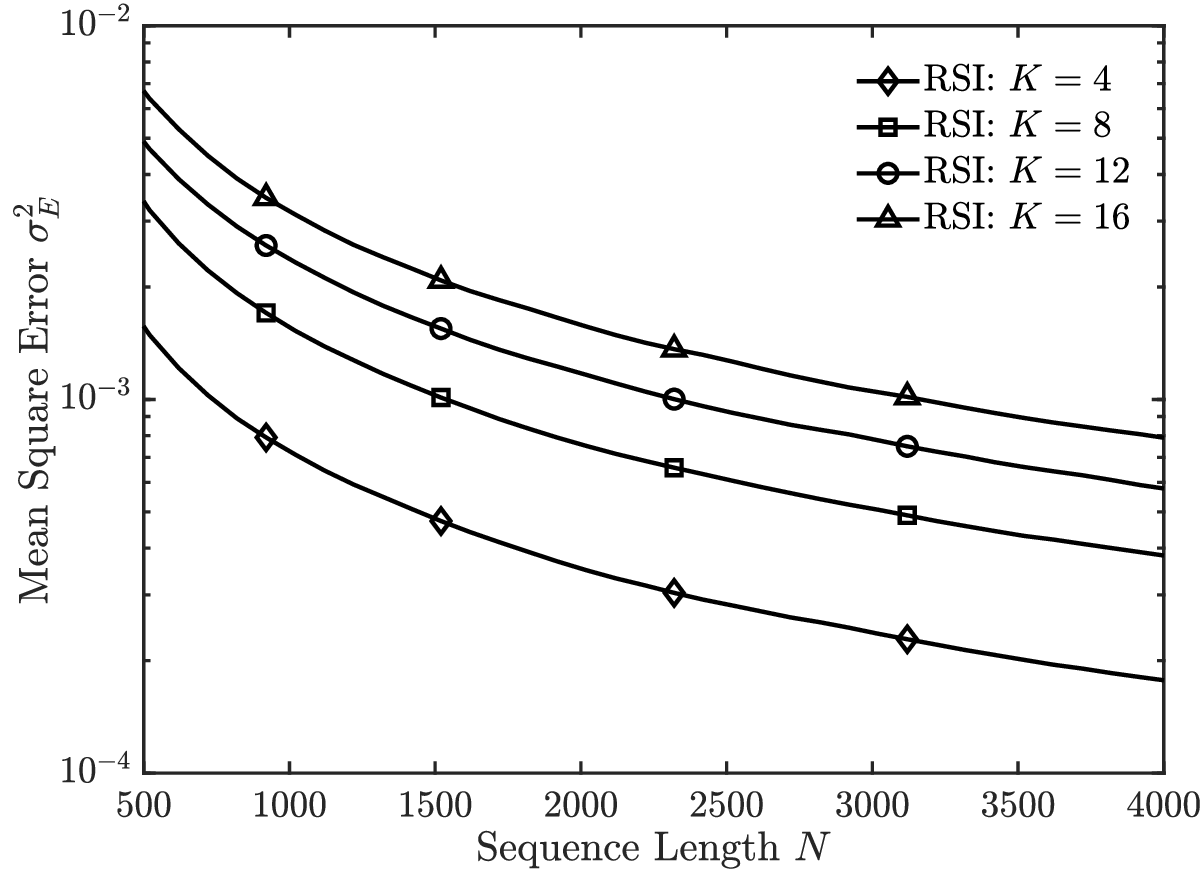}
	\caption{The mean square error $\sigma_E^2$ versus the length of the transmitted symbol sequence $N$.}
	\label{mse_figure}
\end{figure}

First, we investigate the influence of the number of transmitted symbols on the estimation error. Fig. \ref{mse_figure} illustrates the mean square error $\sigma_E^2$ as a function of the transmitted symbol sequence length $N$. In this simulation, the transmitted symbols are BPSK-modulated, i.e., $x_n \in \{+1, -1\}$. As shown in Fig. \ref{mse_figure}, the channel estimation error decreases as $N$ increases for all cases, considering different self-interference channel conditions. Additionally, Fig. \ref{mse_figure} demonstrates that the proposed high-precision channel estimation method performs better in scenarios with simpler self-interference channel environments, corresponding to smaller values of $K$.

\begin{figure}[!t]
	\includegraphics[width=0.48\textwidth]{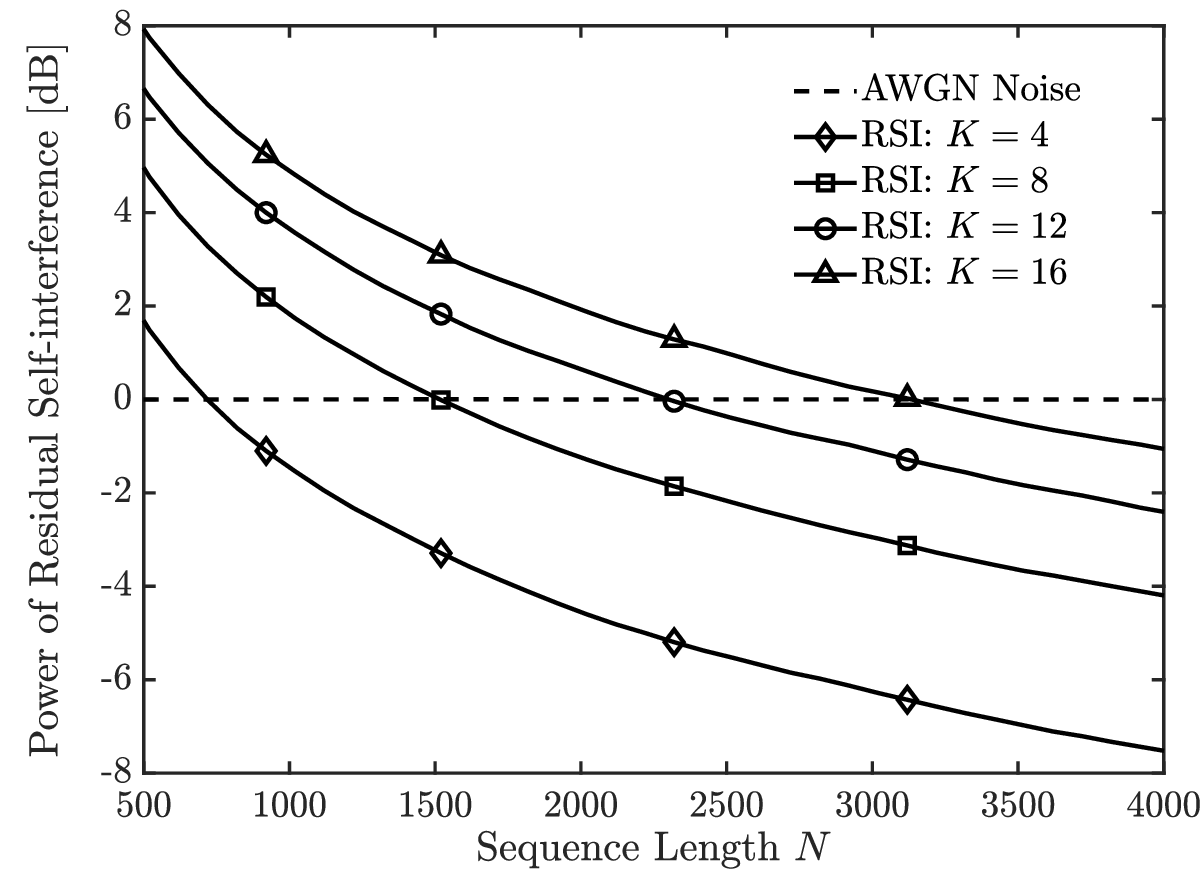}
	\caption{The power of residual self-interference versus the length of the transmitted symbol sequence $N$.}
	\label{sic_figure}
\end{figure}

Based on the high-precision channel estimation, numerical simulations are conducted to evaluate the power of residual self-interference as a function of the transmitted symbol sequence length $N$, which is depicted in Fig. {\ref{sic_figure}}. As shown in Fig. {\ref{sic_figure}}, it can be observed that the residual self-interference undergoes attenuation with increasing values of $N$. Moreover, the magnitude of residual self-interference falls below the noise level when $N$ surpasses a specific threshold, i.e., $N_0=2000$ for the case with $K=20$. Notably, the threshold value $N_0$ depends on the self-interference channel conditions characterized by the value of $K$. Specifically, a larger value of $K$ corresponds to a higher threshold value $N_0$.

\section{Conclusions}
In this paper, we have proposed a novel digital self-interference cancellation method based on high-precision self-interference channel estimation. By utilizing all transmitted symbol for channel coefficients estimation, we have successfully reduced the estimation error. Consequently, the reconstructed self-interference signal has been obtained by convolving the transmitted symbol sequence with the self-interference impulse response. Subsequently, the reconstructed self-interference has been subtracted from the received signal, thereby achieving digital self-interference cancellation. Our numerical results have demonstrated that both the estimation error and residual self-interference signal decrease with an increasing number of transmitted symbols used for channel estimation. This signifies the effectiveness and the improved performance of our proposed method in mitigating self-interference in full duplex communication systems.

\end{document}